\documentclass[prl,groupedaddress,reprint,numerical]{revtex4-1}
\usepackage{amssymb}  
\usepackage{soul}
\usepackage{multirow}
\newcommand{\RNum}[1]{\uppercase\expandafter{\romannumeral #1\relax}}
\usepackage{amsmath}
\usepackage{graphicx}%
\usepackage{dcolumn}%
\usepackage{bm}%
\usepackage{hyperref}
\usepackage[mathscr]{euscript}
\usepackage{color}

\usepackage[normalem]{ulem}

\begin{document}

\title{Fingerprints of angulon instabilities in the spectra of matrix-isolated molecules}
\author{Igor N. Cherepanov}
\email{igor.cherepanov@ist.ac.at}
\author{Mikhail Lemeshko}
\email{mikhail.lemeshko@ist.ac.at}
\affiliation{IST Austria (Institute of Science and Technology Austria), Am Campus 1, 3400 Klosterneuburg, Austria}
\date{\today}

\begin{abstract}
The formation of vortices is usually considered to be the main mechanism of angular momentum disposal in superfluids. Recently, it was predicted that a superfluid can acquire angular momentum via an alternative, microscopic route -- namely, through interaction with rotating impurities, forming so-called `angulon quasiparticles' [Phys.~Rev.~Lett.~\textbf{114}, 203001~(2015)]. The angulon instabilities correspond to transfer of a small number of angular momentum quanta from the impurity to the superfluid, as opposed to vortex instabilities, where angular momentum is quantized in units of $\hbar$ per atom. Furthermore, since conventional impurities (such as molecules) represent three-dimensional (3D) rotors, the  angular momentum transferred is intrinsically 3D as well, as opposed to a merely planar rotation which is inherent to vortices.  Herein we show that the angulon theory can explain the anomalous broadening of the spectroscopic lines observed for CH$_3$ and NH$_3$ molecules in superfluid helium nanodroplets, thereby providing a fingerprint of the emerging angulon instabilities in experiment. 
\end{abstract}

\maketitle
One of the distinct features of the superfluid phase is the formation of vortices -- topological defects carrying quantized angular momentum, which arise if the bulk  of the superfluid rotates faster than some critical angular velocity~\cite{Pitaevskii2016, LeggettQuantLiquids}. Vortex nucleation has been considered to be the main mechanism angular momentum disposal in superfluids~\cite{Pitaevskii2016, PethickSmith, LeggettQuantLiquids, GomezPRL12, Gomez906}. 
Recently, it was predicted that a superfluid can acquire angular momentum via a different, \textit{microscopic} route, which takes effect in the presence of rotating impurities, such as molecules~\cite{ToenniesAngChem04, ToenniesARPC98, Callegari2001, StienkemeierJPB06, Choi2006, SzalewiczIRPC08, VilesovPRL95, WhaleyJCP00}. In particular, it was demonstrated that a rotating impurity immersed in a superfluid forms the `angulon' quasiparticle, which can be thought of as a rigid rotor dressed by a cloud of superfluid excitations carrying angular momentum~\cite{SchmidtLem15, SchmidtLem16,Lemeshko2016, Redchenko16, Li16, Bikash16, Yakaboylu16, Bighin17}.

The angulon theory was able to describe,  in good agreement with experiment, renormalization of rotational constants~\cite{LemeshkoDroplets16, YuliaPhysics17} and laser-induced dynamics~\cite{Shepperson16, Shepperson17} of molecules in superfluid helium nanodroplets. One of the key predictions of the angulon theory are the so-called  `angulon instabilities'~\cite{SchmidtLem15, SchmidtLem16,Lemeshko2016} that occur at some critical value of the molecule-superfluid coupling where the angulon quasiparticle becomes unstable and one or a few quanta of angular momentum are resonantly transferred from the impurity to the superfluid. These instabilities are fundamentally different from the vortex instabilities, associated with the transfer of angular momentum quantised in units of $\hbar$ \textit{per atom} of the superfluid. Furthermore, vortices can be thought of as planar rotors, i.e., the eigenstates of the $\hat L_z$ operator. Angulons, on the other hand, are the eigenstates of the total angular momentum operator, $\hat{\mathbf L}^2$, and  therefore the transferred angular momentum is three-dimensional.  While vortex instabilities have been subject to several experimental studies in the context of superfluid helium \cite{Blaauwgeers2000,Bewley2006,  GomezPRL12, Zmeev2013, Gomez906, Spence2014,Thaler2014,Jones2016}, ultracold quantum gases \cite{Matthews1999,Burger2001,Zwierlein2005,Zwierlein2005,Lin2009,Freilich2010}, and superconductors \cite{Harada1992,Harada1996,Wallraff2003,Guillamon2009}, the transfer of angular momentum to a superfluid via the angulon instabilities has not yet been observed in experiment.

In this Letter we provide  evidence for the emergence of the angulon instabilities in experiments on CH$_3$~\cite{MorrisonJPCA13} and NH$_3$~\cite{Slipchenko2005} molecules trapped in superfluid helium nanodroplets. Spectroscopy of molecules matrix-isolated in $^4$He has been an active area of research during the last two decades~\cite{ToenniesAngChem04, ToenniesARPC98, Callegari2001, StienkemeierJPB06, Choi2006, SzalewiczIRPC08, VilesovPRL95, WhaleyJCP00,MudrichIRPC14}. In general, it is believed  that the superfluid helium environment alters the molecular rovibrational spectra only weakly, the main effect being the renormalization of the molecular moment of inertia~\cite{ToenniesAngChem04}, which is somewhat analogous to the renormalization of effective mass of electrons propagating in crystals~\cite{LemeshkoDroplets16}. It has been shown, however, that superfluid $^4$He leads to homogeneous broadening of some spectroscopic lines. While the inhomogeneous line broadening is known to arise due to the size distribution of the droplets~\cite{ToenniesAngChem04}, the mechanisms of the homogeneous broadening have been under active discussion~\cite{VilesovPRL95,Callegari2001,Slipchenko2005,Scheele2005,MorrisonJPCA13,Ravi2011,Merritt2004, VilesovPRL96,NautaPRL99, NautaJCP00,NautaJCP01,Slenczka2001, ZillichJCP10,VonHaeften2006,ZillichJCP08,ZillichPRB06,ZillichJCP08,ZillichPRL04,ZillichPRB04,ZillichJPCA07,Moore2003,Gutberlet2011,Lehmann2007,Rudolph2007,Skvortsov2009,Pentlehner2010,Hoshina2010,Zhang2014}  and  their convincing microscopic interpretation has been wanting.

\begin{figure}[h!]
\centering
\includegraphics[width=0.50 \textwidth]{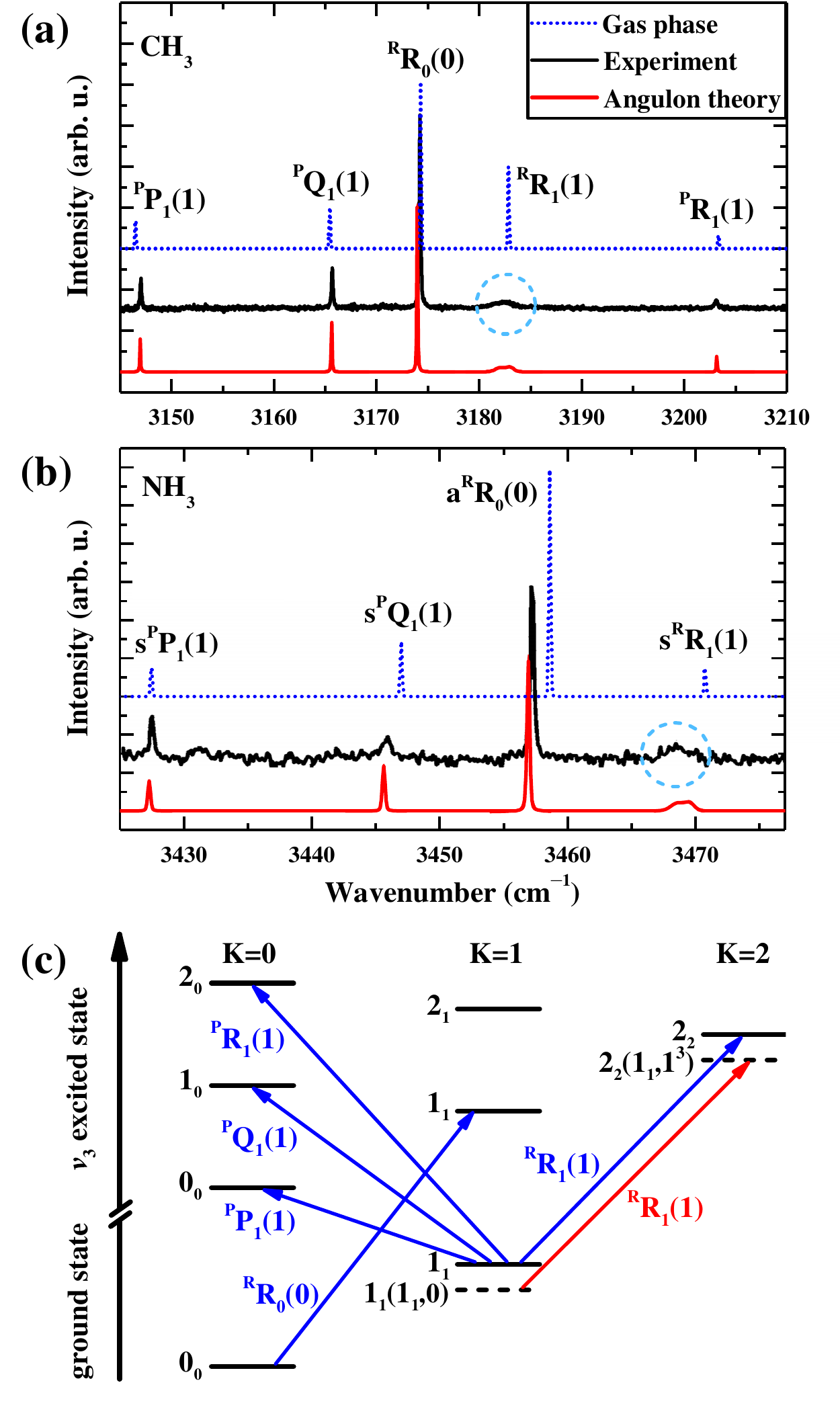}\\[-10pt]
\caption{(a) $\nu_3$ ro-vibrational spectrum of CH$_3$: gase phase simulation (dashed blue) and experiment in superfluid helium nanodroplets (black)  in comparison with the angulon theory (red). (b) Same as (a) for NH$_3$.  (c) Schematics of relevant gas-phase molecular levels (black solid lines) and the corresponding spectroscopic transitions (blue arrows). The case of NH$_3$ involves additional inversion doubling of the levels, which is not shown~\cite{Ruzi2013}. In the presence of helium, the $\mathrm{^RR}_1(1)$ transition (red line) takes place between the angulon states, $1_1 (1_1, 0)$ and $2_2 (1_1,1^3)$ (dashed lines), which results in the line broadening encircled in (a) and (b). Experimental data adapted with permission from Refs.~\cite{MorrisonJPCA13, Slipchenko2005}.}
\label{fig:exp}
\end{figure}

Our aim here is to explain the anomalously large broadening of the $\mathrm{^R R_1 (1)}$ transition, recently observed in in $\nu_{3}$ rovibrational spectra of CH$_3$ in helium droplets~\cite{MorrisonJPCA13}. The experimental spectrum is shown in Fig.~\ref{fig:exp}(a) by the black solid line; Fig.~\ref{fig:exp}(c) provides a schematic illustration of the molecular levels in the gas phase.  One can see that all the spectroscopic lines are left intact by the helium environment, except the $\mathrm{^R R_1 (1)}$ line, which is broadened by  $\sim 50$~GHz~\cite{supplement} compared to the gas-phase simulation (blue dots). In Ref. \cite{MorrisonJPCA13} this feature was qualitatively explained by the coupling between the $2_2$ and $1_1$ molecular levels induced by the $V_{33}(r)$ anisotropic term of the CH$_3$--He potential energy surface (PES)  \cite{supplement}, based on the theory of Ref.~\cite{ZillichJCP10}.   A similar effect was also present in earlier experiments on NH$_3$~\cite{Slipchenko2005}, see Fig.~\ref{fig:exp}(b).  Our goal  is to provide a microscopic description of the  spectra shown using the angulon theory and thereby demonstrate that the broadening is due to an angulon instability, accompanied by a resonant transfer of  $3\hbar$ of angular momentum from the molecule to the superfluid. It is important to note that the angulon quasiparticle theory described below is substantially simpler -- and therefore more transparent -- than  numerical calculations based on  Monte-Carlo algorithms~\cite{ZillichJCP08,ZillichPRB06,ZillichJCP08,ZillichPRL04,ZillichPRB04,ZillichJPCA07,ZillichJCP10,Moroni2004,Moroni2003,Paesani2003,Blinov2004,Miura2007,Skrbic2007,Paesani2005,Paolini2005,Markovskiy2009,Wang2011,Rodriguez-Cantano2013}.

We start by generalizing the angulon Hamiltonian, derived in Refs.~\cite{SchmidtLem15, Lemeshko2016} for linear-rotor molecules, to the case of symmetric tops such as CH$_3$ and NH$_3$~\cite{supplement}:
 \begin{gather} 
\begin{split}
\label{eq:hamil}
& \hat H=B\boldsymbol{\hat{\mathrm{J}}^2}+(C-B)\hat{J}^{\prime2}_z+\sum_{q \lambda \mu} \omega (q) \hat{b}^{\dag}_{q \lambda \mu} \hat{b}_{q \lambda \mu}+ \\ + \alpha \sum_{q \lambda \mu \xi}  &v_{\lambda \xi}(q) \bigg(\hat{b}^\dagger_{q \lambda \mu} [D^{\lambda}_{\mu \xi} (\hat{\Omega})+(-1)^{\xi}D^{\lambda}_{\mu -\xi} (\hat{\Omega})]+  \text{h.c.} \bigg)
\end{split}
\end{gather}
where we introduced the notation $\sum_q \equiv \int dq$ and set $\hbar \equiv 1$. The first two terms of Eq.~\eqref{eq:hamil} correspond to the kinetic energy of a symmetric-top impurity, with $\mathbf{\hat{J}}$ and $\mathbf{\hat{J'}}$ the angular momentum operators acting in the laboratory and impurity frames, respectively~\cite{Herzberg1945, BernathBook, Bunker2006, LevebvreBrionField2}. $B$ and $C$ are rotational constants determined by the corresponding moments of inertia as $B=\frac{1}{2I_{x'}}=\frac{1}{2I_{y'}}$ and $C=\frac{1}{2I_{z'}}$. The energies of the free impurity states are given by $E_{JK}=BJ(J+1)+(C-B)K^2$, and correspond to $(2J+1)$--fold degenerate states, $\vert JMK \rangle$. Here $J$ is the angular momentum of the molecule, $M$ gives its projection on the $z$-axis of the laboratory frame, and $K$ gives its projection on the $z'$-axis of the molecular frame. 

The third term of the Hamiltonian represents the kinetic energy of the bosons in the superfluid, as given by the dispersion relation, $\omega (q)$. Here the boson creation and annihilation operators, $\hat{b}^{\dag}_{q \lambda \mu}$ and $\hat{b}_{q \lambda \mu}$, are expressed in the angular momentum basis, where $q=\vert \boldsymbol{\mathrm{q}} \vert$ labels the boson's linear momentum, $\lambda$ is the angular momentum, and $\mu$ is the angular momentum projection onto the $z$-axis, see Ref.~\cite{Lemeshko2016} for details.

 The last term of Eq.~\eqref{eq:hamil} defines the interactions between the molecular impurity and the superfluid, where we have introduced an auxiliary parameter $\alpha$, which for comparison with experiment will be set to $\alpha \equiv 1$. $D^{\lambda}_{\mu \xi} (\hat{\Omega})$ are Wigner $D$-matrices, and $\hat{\Omega} \equiv (\hat{\theta}, \hat{\phi}, \hat{\gamma})$ are the angle operators defining the orientation of the molecular axis in the laboratory frame. It is important to note that Eq.~\eqref{eq:hamil} becomes quantitatively accurate for a symmetric-top molecule immersed in a weakly-interacting Bose-Einstein Condensate~\cite{SchmidtLem15, Lemeshko2016}. It has been demonstrated, however, that one can develop a phenomenological theory based on the angulon Hamiltonian that describes rotations of molecules in superfluid $^4$He in good agreement with experiment~\cite{LemeshkoDroplets16, Shepperson16}. Here we pursue a similar   route, i.e., we fix the interaction parameters, $v_{\lambda \xi}(q)$, based on \textit{ab inito} PES's~\cite{supplement,Meyer1986,Hodges2001,Dagdigian2011,Suarez2011,Green1976,Green1980} in such a way that the depth of the trapping potential (mean-field energy shift) for the molecule is reproduced~\cite{ToenniesAngChem04}.
 
Let us proceed with calculating the spectrum of a symmetric-top impurity in $^4$He in the weakly-interacting regime, applicable to both CH$_3$ and NH$_3$~\cite{LemeshkoDroplets16}. We start from constructing a variational ansatz based on single-boson excitations, analogous to that used in Ref.~\cite{SchmidtLem15} for linear molecules:
\begin{gather} 
\begin{split}
\label{eq:chevy}
\vert \psi_{LMk_0} \rangle= Z^{1/2}_{LMk_0}& \vert 0 \rangle \vert LMk_0 \rangle +\\ &+\sum_{\begin{subarray}kq\lambda \mu \\ jmk \end{subarray}} \beta_{\lambda j k} (q) C^{LM}_{jm, \lambda \mu} \hat{b}^{\dag}_{q \lambda \mu} \vert 0 \rangle \vert jmk \rangle
\end{split}
\end{gather}
Here $\vert 0 \rangle$ is the vacuum of bosonic excitations, and $Z^{1/2}_{LMk_0}$ and $\beta_{\lambda j k} (q)$ are the variational parameters obeying the normalization condition, $Z_{LMk_0}=1-\sum_{q\lambda jk} \vert \beta_{\lambda jk}(q) \vert ^2$. The coefficient $Z_{LMk_0}$ is the so-called quasiparticle weight~\cite{LL9, AltlandSimons}, i.e., the overlap between the dressed angulon state, $\vert \psi_{LMk_0} \rangle$, and the free molecular state, $\vert LMk_0 \rangle \vert 0 \rangle$.

The angulon state~\eqref{eq:chevy} is an eigenstate of the total angular momentum operators, $\mathbf{\hat L}^2$ and $\hat{L}_z$, which correspond to good quantum numbers $L$ and $M$. In the absence of external fields, the quantum number $M$ is irrelevant and will be omitted hereafter.  In addition, we introduce approximate quantum numbers $j$ and $k$, describing the angular momentum of the molecule and its projection on the molecular axis $z'$ ($k = k_0$ for $j=L$), and $\lambda$ giving angular momentum of the excited boson. The idea of approximate quantum numbers in the present context is analogous to (and inspired by) Hund's cases of molecular spectroscopy~\cite{LevebvreBrionField2}. As a result, we can label the angulon states  as $L_{k_0} (j_k, N_{\lambda}$$^{\lambda})$, where $N_{\lambda}$ gives the number of phonons in a state with angular momentum $\lambda$. The ansatz of Eq.~\eqref{eq:chevy} restricts the possible values of  $N_{\lambda}$ to 0 or 1.

After the minimization of the energy, $E = \langle \psi_{LMk_0} \vert  \hat H \vert \psi_{LMk_0}  \rangle/ \langle \psi_{LMk_0} \vert \psi_{LMk_0}  \rangle$, with respect to $Z^{1/2 \ast}_{LMk_0}$ and $\beta_{\lambda j k}^\ast (q)$, we arrive to the Dyson-like equation~\cite{SchmidtLem15, Lemeshko2016}:
\begin{equation}
\label{eq:Dyson}
E=BL(L+1)+(C-B)k_0^2-\Sigma_{Lk_0} (E)
\end{equation}
Here $\Sigma_{Lk_0} (E)$ is the angulon self-energy containing all the information about the molecule-helium interaction:
\begin{gather} 
\label{eq:sigma}
\begin{split}
&\Sigma_{Lk_0} (E)=\sum_{q \lambda j k \xi \xi'} \frac{v_{\lambda \xi} (q) v_{\lambda \xi'} (q) }{Bj(j+1) +(C-B)k^2 -E+\omega (q)} \times
\\\times (C&^{jk}_{LK, \lambda \xi}+(-1)^{\xi } C^{jk}_{LK, \lambda -\xi})(C^{jk}_{LK, \lambda \xi'}+(-1)^{\xi'} C^{jk}_{LK, \lambda -\xi'})
\end{split}
\end{gather}
In the limit of $k=0, K=0$, $\xi=0$, and $\xi'=0$,  Eqs.~\eqref{eq:Dyson} and \eqref{eq:sigma} reduce to the equations derived in Ref.~\cite{SchmidtLem15} for a linear molecule.

\par
Within the electric dipole approximation, the angulon excitation spectrum  is given by the following expression:
\begin{gather} 
\begin{split}
\label{eq:spectrum}
S^{v' v}_{LMk_0}(E)=\vert \langle v' \vert \langle &\psi_{L'M'k_0'} \vert \boldsymbol{\hat{\mu}} \vert \psi_{LMk_0} \rangle \vert v \rangle \vert ^2 \times \\ \times &\frac{\mathrm{Im}\Sigma_{L'k_0'}(E)}{(\gamma_{L'k_0'}(E)-E)^2+[\mathrm{Im}\Sigma_{L'k_0'}(E)]^2}
\end{split}
\end{gather}
where $\gamma_{L'k_0'}(E)=BL'(L'+1)+(C-B)k_0'^2-\mathrm{Re}\Sigma_{L'k_0'}$, $\boldsymbol{\hat{\mu}}$ is the dipole moment operator, and  $\vert v \rangle$ and $\vert v' \rangle$ label the initial and final vibrational states. We assume that only one initial state, $\vert \psi_{LMk_0} \rangle \vert v \rangle$, is populated and the optical transition occurs to all excited states, $\vert \psi_{L'M'k_0'} \rangle \vert v' \rangle$, in accordance with the selection rules determined by the electric dipole matrix elements. As can be seen, the imaginary part of the self-energy, $\mathrm{Im}\Sigma_{L'k_0'}(E)$, gives the width of the spectral lines.  It is important to note that the angulon Hamiltonian~\eqref{eq:hamil} describes solely the rotational motion and does not explicitly take into account any effects related to molecular vibrations. While the vibrational corrections due to helium are relatively small~\cite{Paesani2002,ToenniesAngChem04,Rudolph2007,ZillichJCP10}, we have included them into our model for a more accurate comparison with experiment~\cite{supplement}.

Let us compare  the prediction of the angulon theory  with the experimental data of Fig.~\ref{fig:exp}(a),(b). In order to obtain quantitative results, we need to fix the model parameters. For the $\nu_{3}$  vibrational band of CH$_3$  the rotational constants are   $B=9.47111(2) ~\mathrm{cm^{-1}}$ and $C=4.70174(3) ~\mathrm{cm^{-1}}$~\cite{Davis1997};  for NH$_3$, $B=9.76647(17) ~\mathrm{cm^{-1}}$ and  $C=6.23370(21) ~\mathrm{cm^{-1}}$~\cite{Abdullah1989,Bach2002}.  For $\omega (q)$ we substitute the empirical dispersion relation~\cite{DonnellyHe98}.   The coupling  constants, $ v_{\lambda \xi}(q)$, can be derived from the Fourier transforms of the spherical components of the PES~\cite{SchmidtLem15, supplement},
\begin{equation}
 v_{\lambda \xi}(q) = \sqrt{\frac{nq^4}{\pi m \omega(q)}}(1+\delta_{\xi 0})^{-1} \int dr r^2 f_{\lambda \xi}(r) j_{\lambda} (r q),
\end{equation}
 where $m$ is the mass of a helium atom,   $ j_{\lambda} (rq)$ are the spherical Bessel functions, and $f_{\lambda \xi}(r) $ determines the components of the spherical harmonics expansion of the  molecule-helium potentials \cite{Meyer1986,Hodges2001,Dagdigian2011,Suarez2011}. In order to derive the simplest possible model, we take into account only the isotropic term, $\lambda = \xi = 0$, as well as the leading anisotropic term, $\lambda = \xi = 3$.  It has been previously shown~\cite{LemeshkoDroplets16, Shepperson16} that the effects of helium can be parametrized by a few characteristic properties of the molecule-helium potential, such as the PES anisotropy and the depth of its minima, which renders the fine details of the PES irrelevant. Therefore, in order to further simplify the model, we choose effective potentials characterized by the Gaussian form-factors, $f_{\lambda \xi}(r)=u_{\lambda \xi} (2\pi)^{-3/2} e^{-\frac{r^2}{2r_{\lambda\xi}^2}}$, such that their magnitude, $u_{\lambda \xi}$, and range, $r_{\lambda\xi}$, reproduce known properties of the molecule-helium interaction. In particular, we set $r_{00}=r_{33}=3.45~\mathrm{\AA}$ ($r_{00}=r_{33}=3.22 ~\mathrm{\AA}$), to the position of the global minimum of the CH$_3-$He~\cite{Dagdigian2011} (NH$_3-$He~\cite{Suarez2011}) PES. The magnitude of the isotropic potential, $u_{00}= 23.2~B$ for CH$_3$ and  $u_{00}= 26.0~B$ for NH$_3$, was chosen so as to reproduce the mean-field shift (`trapping depth' or `impurity chemical potential') of $40 ~\mathrm{cm^{-1}}$, typical for small molecules dissolved in helium nanodroplets \cite{ToenniesAngChem04}. Finally, the anisotropy ratio, $u_{33}/u_{00}=0.22$ for CH$_3$ and $u_{33}/u_{00}=0.25$ for NH$_3$, was chosen to reproduce the ratio of the areas under the corresponding \textit{ab initio} PES components~\cite{supplement}.

\begin{figure}[ht!]
\centering
\includegraphics[scale=0.55]{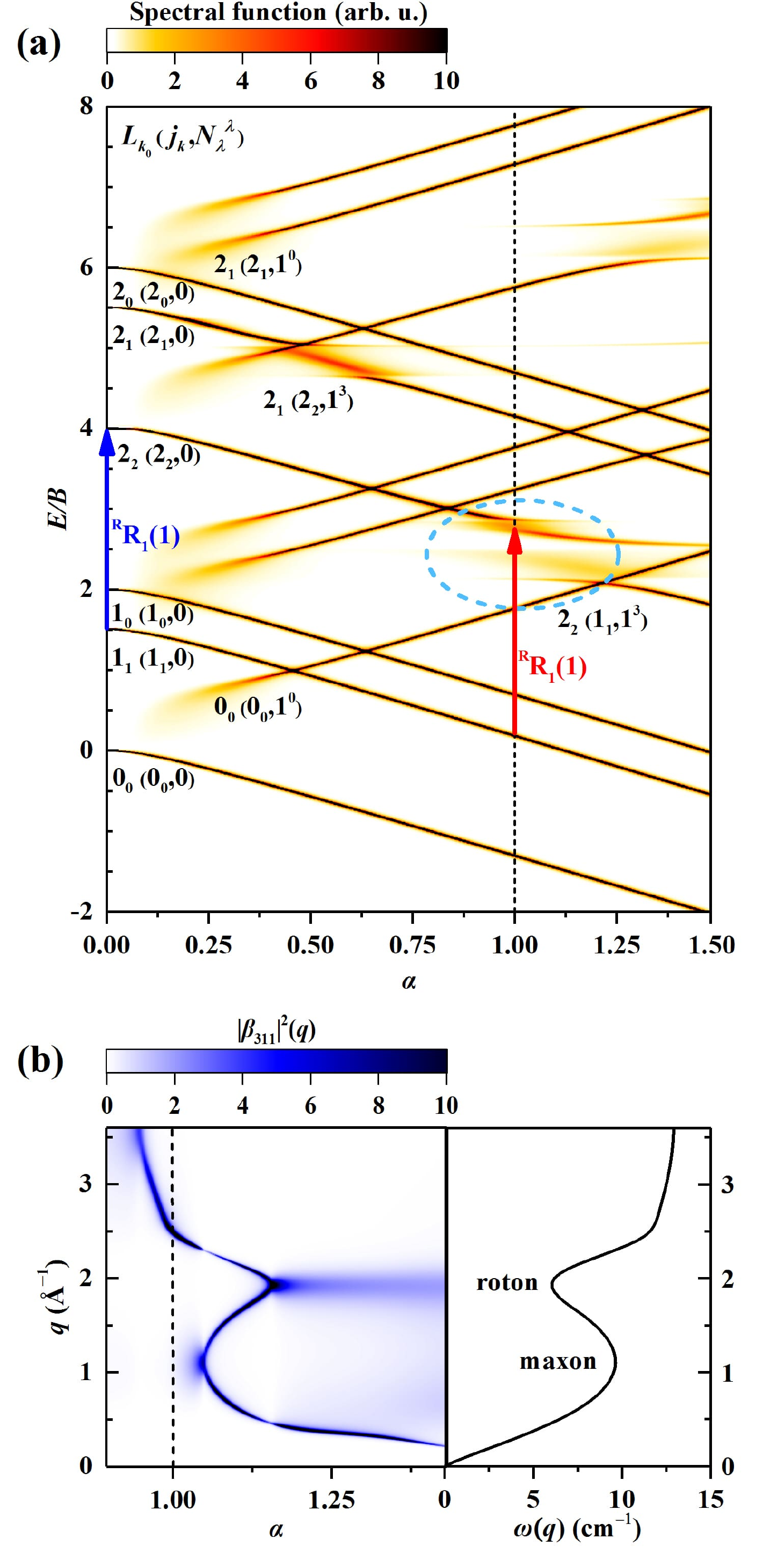}
\caption{(a) Angulon spectral function for CH$_3$ as a function of the dimensionless energy $E/B$ and interaction parameter $\alpha$. The spectrum of Fig.~\ref{fig:exp}(a) corresponds to $\alpha\equiv1$ (dashed vertical line). The broadening of the $\mathrm{^RR_1(1)}$  line  occurs due to the angulon instability (encircled). (b) Phonon density $\vert \beta_{311}(q) \vert ^2$ in the vicinity of the angulon instability (left) and the experimentally measured dispersion relation of bulk $^4$He~\cite{DonnellyHe98} (right). The phonon excitation takes place in the region of wavevectors $q \sim 2.5$~\AA. }
\label{fig:spectral}
\end{figure}

Red lines in Fig.~\ref{fig:exp}(a), (b) show the results of the angulon theory from Eqs.~\eqref{eq:sigma}--\eqref{eq:spectrum}, with $\alpha \equiv 1$. One can see that the angulon theory is in a good agreement with experiment for all  the spectroscopic lines considered. In particular, for the broadened $\mathrm{^RR}_1(1)$ line, we obtain  the linewidth of 50 GHz for CH$_3$ and 47 GHz for NH$_3$, which is close to the experimental values of 57~GHz and 50~GHz, respectively.
In order to gain insight into the origin of the line broadening, let us study how the angulon spectral function~\cite{SchmidtLem15, LL9, AltlandSimons}  changes with the molecule-helium interaction strength. The spectral function can be obtained from Eq.~\eqref{eq:spectrum} by setting $\langle v' \vert \langle \psi_{L'M'k_0'} \vert \boldsymbol{\hat{\mu}} \vert \psi_{LMk_0} \rangle \vert v \rangle  \equiv 1 $, which corresponds to neglecting all the spectroscopic selection rules. Fig.~\ref{fig:spectral}(a) shows the spectral function for the parameters of the CH$_3$ molecule listed above, as a function of  energy, $E/B$, and the molecule-helium interaction parameter, $\alpha$.  The corresponding spectral function for NH$_3$ looks qualitatively similar.

The limit of $\alpha \to 0$ corresponds to the states of a free molecule, shown in Fig.~\ref{fig:exp}(c). For finite $\alpha$, however, the angulon levels develop an additional fine structure, which was discussed in detail in  Refs.~\cite{SchmidtLem15, SchmidtLem16, Lemeshko2016}. Of a particular interest is the region in the vicinity of $\alpha=1$, where the so-called `angulon instability' occurs. In this region, 
the state with total angular momentum $L=2$ (which is a good quantum number) changes its composition: for $\alpha \lesssim 0.8$, the angulon state corresponds to $2_2 (2_2, 0 )$, i.e., it is dominated by the molecular state $2_2$.  In the region of $0.8 \lesssim \alpha \lesssim 1.3$, the $L=2$ angulon state crosses the phonon continuum attached to the $j=1$ molecular state, which results in the phonon excitation. The collective state in the instability region is $2_2 (1_1, 1^3)$. That is, while the total angular momentum $L=2$ is conserved,  it is shared between the molecule and the superfluid due to the molecule-helium interactions. Effectively, the molecule finds itself in the $1_1$ state, which is accompanied by a creation of one phonon with angular momentum $\lambda=3$. Fig.~\ref{fig:spectral}~(b) shows the  phonon density, $|\beta_{3 1 1} (q)|^2$, in the vicinity of the instability, which is dominated by short-wavelength excitations with $q \sim 2.5$~\AA~located in the `beyond the roton' region \cite{Azuah2013}.

It is important to note that the qualitative discussion of Ref.~\cite{MorrisonJPCA13} attributed the broadening to excitation of at least two phonons,  since the splitting between the $2_2$ and $1_1$ states ($\sim 21~\mathrm{cm^{-1}}$) exceeds the maximum energy of an elementary excitation in superfluid $^4$He ($\sim 14~\mathrm{cm^{-1}}$~\cite{Manousakis1986}). The results presented above demonstrate that the broadening can be explained as a one-phonon transition  between two many-particle states, $2_2 (2_2, 0 )$ and $2_2 (1_1, 1^3)$. In other words, in the presence of the superfluid, the level structure of the dressed molecule changes, and the energy conservation arguments have to be modified accordingly. 
 
Thus, we have generalized the angulon theory to the case of light symmetric-top molecules and demonstrated that   angulon instabilities predicted in Refs.~\cite{SchmidtLem15, SchmidtLem16} have in fact been observed in the spectra of CH$_3$ and NH$_3$ immersed in superfluid helium nanodroplets. This paves the way to studying the decay of angulon quasiparticles and other microscopic mechanisms of the angular momentum transfer in experiments on quantum liquids, with possible applications to phonon quantum electrodynamics~\cite{SoykalPRL11}.  Furthermore, the angulon instabilities have been predicted to lead to anomalous screening of quantum impurities~\cite{Yakaboylu16} as well as to the emergence of non-abelian magnetic monopoles~{\cite{Yakaboylu2017}}, which opens the door for the study of exotic physical phenomena in helium droplet experiments. Future measurements on isotopologues, such as CD$_3$ or ND$_3$, would allow the variation of the  molecular rotational constants without altering the molecule-helium interactions,  thereby providing an additional test of the model.  It would be of great interest to perform quench experiments involving short laser pulse excitations~\cite{PentlehnerPRL13, Shepperson16, Shepperson17} of CH$_3$ and NH$_3$ in helium nanodroplets aiming to observe the dynamical emergence of the angulon instability.
 
\begin{acknowledgments}
We thank Richard Schmidt for comments on the manuscript and Gary Douberly for insightful discussions and providing the experimental data from Ref.~\cite{MorrisonJPCA13}. This work was supported by the Austrian Science Fund (FWF), project Nr. P29902-N27 and by the European Union’s Horizon 2020 research and innovation programme under the Marie Sklodowska-Curie Grant Agreement No. 665385.
\end{acknowledgments}

\bibliography{Ref} 

\newpage
\clearpage

\vspace{0cm}

\renewcommand{\thepage}{S\arabic{page}}  
\renewcommand{\thesection}{S\arabic{section}}   
\renewcommand{\thetable}{S\arabic{table}}   
\renewcommand{\thefigure}{S\arabic{figure}}
\renewcommand{\theequation}{S\arabic{equation}}

\setcounter{figure}{0} 
\setcounter{table}{0} 
\setcounter{equation}{0} 
\setcounter{page}{1} 

\onecolumngrid
\section{Supplemental Material}

\subsection{Derivation of the angulon hamiltonian}
\begin{figure}[h!]
\centering
\includegraphics[width=0.6\textwidth]{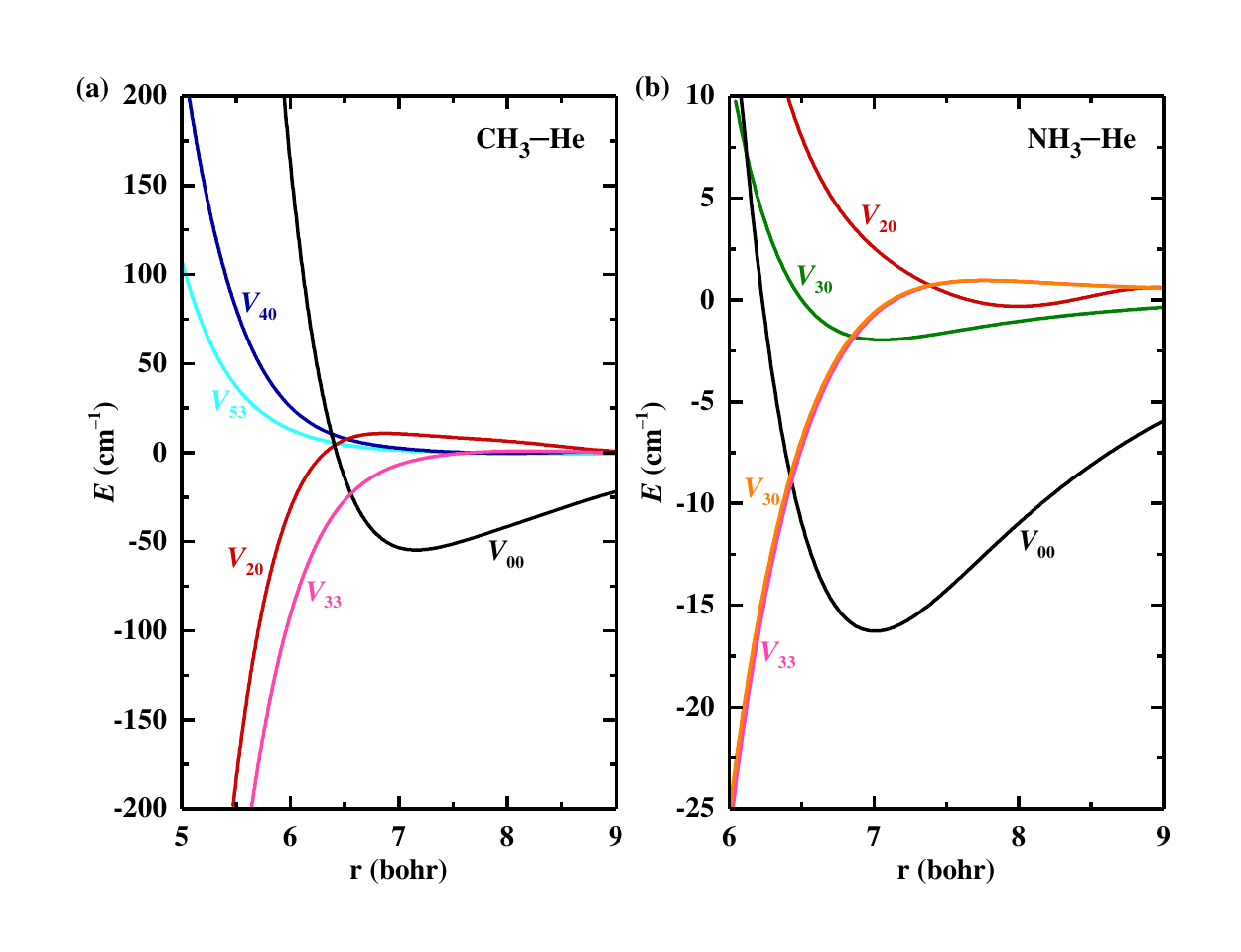}
\caption{(a) Spherical components, $ V_{\lambda \xi}(r)$, of the CH$_3-$He PES~\cite{Dagdigian2011}. (b) Same as (a) for NH$_3-$He~\cite{Suarez2011}.}
\label{fig:curves}
\end{figure}

The Hamiltonian for a rotating molecule interacting with a bath of bosons has the following structure: $\hat H= \hat H_\text{mol}+  \hat H_\text{bos}+  \hat H_\text{mol-bos}$. The present derivation for a symmetric-top impurity is analogous to the linear impurity case described in detail in Refs.~\cite{Lemeshko2016, SchmidtLem15, SchmidtLem16}.  For a symmetric-top impurity, the anisotropic molecule-helium potential is expanded in spherical harmonics as follows~\cite{Green1976,Green1980}:
\begin{equation}
V_\text{mol-He}(r, \theta, \phi)= \sum_{\lambda \xi}  V_{\lambda \xi} (r)(1+\delta_{\xi 0})^{-1} [ Y_{\lambda \xi} (\theta, \phi)+ Y_{\lambda \xi}^* (\theta, \phi)]
\end{equation}
Here $(r, \theta, \phi)$ describe the position of the helium atom with respect to the center of mass of the molecule in the molecular (body-fixed) coordinate system. Spherical components, $V_{\lambda \xi}(r)$, of the PES for the CH$_3-$He and NH$_3-$He complexes \cite{Dagdigian2011,Suarez2011} are shown in Fig.~\ref{fig:curves}.
\par
 The pairwise interaction potential determines the explicit form for the last term of the Hamiltonian~\eqref{eq:hamil}, which contains the coupling constants $v_{\lambda \xi}(q)$, defined as:
\begin{equation}
 v_{\lambda \xi}(q) = \sqrt{\frac{2nq^2 \epsilon (q)}{\pi \omega (q)}}(1+\delta_{\xi 0})^{-1} \int dr r^2 f_{\lambda \xi}(r) j_{\lambda} (rq)
\end{equation}
Here $\epsilon (q)=\frac{q^2}{2m}$ is the kinetic energy of a boson of mass $m$, and $ j_{\lambda} (rq)$ are spherical Bessel functions. We choose model interaction potentials characterized by the Gaussian form-factors, $f_{\lambda \xi}(r)=u_{\lambda \xi}(2\pi)^{-3/2}e^{-\frac{r^2}{2r_{\lambda\xi}^2}}$, with parameters $r_{\lambda \xi}$ corresponding to the global minimum of the   CH$_3$--He (NH$_3$--He) PES. The isotropic component, $u_{00}$, was chosen such that it  reproduces the mean-field shift (`trapping potential') of 40 $\mathrm{cm^{-1}}$, typical for small molecules in helium nanodroplets~\cite{ToenniesAngChem04}. The mean-field shift can be expressed in terms of $f_{00} (r)$ as follows \cite{Lemeshko2016}:
\begin{equation}
E_\text{mf}= \sqrt{4 \pi} n \int r^2 f_{00}(r)dr
\end{equation}
The ratio of $u_{33}/u_{00}$ was fixed to satisfy   the following condition:
\begin{equation}
  \frac{\int^{\infty}_0 dr r^2 f_{33}(r) j_{3} (r q)}{\int^{\infty}_0 dr r^2 f_{00}(r) j_{0} (r q)} = \frac{\int_{r_c}^{\infty} dr r^2  V_{33}(r) j_{3} (r q)}{\int_{r_c}^{\infty} dr r^2  V_{00}(r) j_{0} (r q)},
  \end{equation}
where $V_{\lambda \xi}(r)$ are the spherical components of the \textit{ab initio}  PES. The cut-off distance, $r_c$, was set to the classical turning point for a collision at the temperature inside a helium droplet, i.e. such that $V_{00}(r_c)=k_B \times 0.4~\mathrm{K}$, with $k_B$ the Boltzmann constant~\cite{ToenniesAngChem04}.
\subsection{The Dyson equation}

Minimization of energy, $E = \langle \psi_{LMk_0} \vert  \hat H \vert \psi_{LMk_0}  \rangle/ \langle \psi_{LMk_0} \vert \psi_{LMk_0}  \rangle$, with respect to $Z^{1/2 \ast}_{LMk_0}$ and $\beta_{\lambda j k}^\ast (q)$ leads to the Dyson-like equation~\cite{Lemeshko2016}:
\begin{equation}
\label{eq:DysonSM}
[G^\text{ang}_{Lk_0} (E)]^{-1}=[G^{0}_{Lk_0} (E)]^{-1}-\Sigma_{Lk_0}(E)=0,
\end{equation}
where
\begin{equation}
 G^{0}_{Lk_0} (E)=\frac{1}{BL(L+1)+(C-B)k_0^2-E}
\end{equation}
is the Green's function of the unperturbed molecule and $G^\text{ang}_{Lk_0} (E)$ is the angulon Green's function. The energy can be found self-consistently, as a set of solutions to  Eq.~\eqref{eq:DysonSM} for a given total angular momentum $L$, which is the conserved quantity of the problem. Alternatively, one can reveal stable and meta-stable states of the system by  calculating the spectral function~\cite{LL9, AltlandSimons, Lemeshko2016}:
\begin{equation}
\label{s:epec}
\mathcal{A}_{Lk_0}=\mathrm{Im}[G^\text{ang}_{Lk_0} (E+i0^+)]
\end{equation}
The spectral function~\eqref{s:epec} corresponds to Eq.~\eqref{eq:spectrum} with $\langle v' \vert \langle \psi_{L'M'k_0'} \vert \boldsymbol{\hat{\mu}} \vert \psi_{LMk_0} \rangle \vert v \rangle  \equiv 1 $.

\subsection{Matrix elements for spectroscopic transitions}

\newcommand{\tj}[6]{ \begin{pmatrix}
  #1 & #2 & #3 \\
  #4 & #5 & #6 
\end{pmatrix}}

Within the electric dipole approximation, the probability of a perpendicular optical transition between two  angulon states, $\vert \psi_{LMk_0}\rangle \vert v \rangle $ and $\vert \psi_{L'M'k'_0}\rangle \vert v' \rangle $, is given by:
\begin{equation}
I_{LMk_0,v}^{L'M'k'_0,v'} \sim \frac{g'}{g} \vert \langle v' \vert \langle \psi_{L'M'k'_0}\vert  \boldsymbol{ \mathrm{\hat{\mu}}} \vert \psi_{LMk_0}\rangle \vert v \rangle \vert^2
\end{equation}
where $g$ and $g'$ give the degeneracies of the $\vert \psi_{LMk_0}\rangle \vert v \rangle $ and $\vert \psi_{L'M'k'_0}\rangle \vert v' \rangle $ states, respectively, $\boldsymbol{ \mathrm{\hat{\mu}}}$ is the dipole moment operator.
Substituting the angulon wavefunctions from Eq.~\eqref{eq:chevy}, for a perpendicular optical transition, we obtain~\cite{Bunker2006}:
\begin{equation}
I_{LMk_0,v}^{L'M'k'_0,v'} \sim \frac{g'}{g} \vert \langle v'\vert \boldsymbol{\mathrm{\hat{\mu}}} \vert v \rangle \vert ^2 \left[ Z^{1/2*}_{LMk_0} Z^{1/2}_{L'M'k'_0}M_{LMk_0}^{L'M'k'_0}+\sum_{\begin{subarray} q~q\lambda \mu \\~jmk \\ j'm'k'\end{subarray}}  \beta^*_{\lambda jk} (q) \beta_{\lambda j'k'} (q) C^{LM}_{jm,\lambda \mu} C^{L'M'}_{j'm',\lambda \mu}M^{j'm'k'}_{jmk} \right]^2
\end{equation}
where $M^{jmk}_{j'm'k'}$ is a rotational matrix element for the transition between the molecular states $\vert jmk \rangle$ and $\vert j'm'k' \rangle$ \cite{Bunker2006}.
\begin{equation}
M^{jmk}_{j'm'k'}=\sqrt{\frac{2j+1}{8\pi^2}}\sqrt{\frac{2j'+1}{8\pi^2}}\Bigg[ \tj{1}{j}{j'}{0}{m}{-m'} \tj{-1}{j}{j'}{0}{k}{-k'} + \tj{1}{j}{j'}{0}{m}{-m'} \tj{1}{j}{j'}{0}{k}{-k'} \Bigg ]
\end{equation}

\subsection{Corrections to the angulon energy}
We corrected the energies of angulons states for the vibrational shift in He droplets, inversion splitting of NH$_3$ and Coriolis coupling. The shift of the vibrational frequency and ground-state inversion splitting of NH$_3$ were set to their empirical values: $\delta\nu=\nu_\text{He}-\nu_\text{gas}=0.08~ \mathrm{cm^{-1}}$ for CH$_3$ \cite{MorrisonJPCA13},  $\delta\nu=-0.5~\mathrm{cm^{-1}}$, $\Delta^\text{inv}_0=0.8~ \mathrm{cm^{-1}}$  for NH$_3$ \cite{Slipchenko2005}. The  rotation-vibration Coriolis coupling is given by the following matrix element \cite{Bunker2006}:
\begin{equation}
\langle v \vert \langle \psi_{Lk_0} \vert - 2\zeta C \hat{\pi}'_z \hat{J'_z} \vert \psi_{Lk_0} \rangle \vert v\rangle = -2\zeta C l \bigg (Z^\ast_{Lk_0}k_0+\sum_{qjk} \vert \beta_{\lambda jk}(q) \vert^2 k\bigg)
\end{equation}
where $\vert \psi_{Lk_0} \rangle$ is the angulon state of Eq.~\eqref{eq:chevy}, $\zeta$ is the constant parametrizing the Coriolis coupling, $\hat{\pi}'_z$ is the vibrational angular momentum operator with respect to the symmetric top axis with eigenvalues $l$. For a given vibrational state $\vert v \rangle$, the vibrational angular momentum $\vert l\vert = v,v-2,\ldots,1$ or $0 $. For the ground vibrational state $l=0$, and $\vert l \vert=1$ for the excited state under consideration, $\vert \nu_3=1 \rangle$. We used the Coriolis constants $\zeta C= 0.35$~cm$^{-1}$ for CH$_3$~\cite{Davis1997}, and $\zeta C = 0.29$~cm$^{-1}$ for NH$_3$~\cite{Abdullah1989}.

\subsection{Comparison to experiment}

Table~\ref{tab:parameters} lists the spectral characteristics of the experimental lines shown in Fig.~\ref{fig:exp}(a), (b), as well as the results of the angulon theory. One can see that the angulon theory is able to reproduce the width of the $^\mathrm{R}$R$_1$(1) line, which is approximately one order of magnitude broader compared to other transitions.  The model tends to underestimate the line broadening by a few GHz, which we attribute to the fact that only single-phonon excitations are included into the ansatz~\eqref{eq:chevy}. Using more involved, diagrammatic approaches~\cite{Bighin17} to the Hamiltonian~\eqref{eq:hamil} is expected to further improve the agreement.

\begin{table}[h!]
\centering
\caption{Comparison of the $\nu_{3}$ spectral line frequencies, $\nu$ (in $\mathrm{cm^{-1}}$), and widths, $\Gamma$ (in GHz),  obtained using the angulon model with the experimental ones. Spectral lines are labeled as $^{\Delta K} \Delta L_{K''} (L'')$, where the initial state is marked with a double prime.  We omit the $\boldsymbol{\mathrm{s}}$ and $\boldsymbol{\mathrm{a}}$ indices labelling the inversion splitting in NH$_3$~\cite{Ruzi2013}. \vspace{0.2cm}}\label{tab:parameters}
\setlength{\extrarowheight}{.2em}
\begin{tabular}{|c|c|c|c|c|c|c|c|c|}
\hline
\multirow{3}{*}{Line} & \multicolumn{4}{c|}{CH$_3$}                                            & \multicolumn{4}{c|}{NH$_3$}                                                \\ \cline{2-9} 
                      & \multicolumn{2}{c|}{Angulon theory} & \multicolumn{2}{c|}{Experiment~\cite {MorrisonJPCA13}} & \multicolumn{2}{c|}{Angulon theory} & \multicolumn{2}{c|}{Experiment \cite{Slipchenko2005}} \\ \cline{2-9} 
                      &       $\nu$            &          $\Gamma$       &       $\nu$         &         $\Gamma$        &      $\nu$                &      $\Gamma$             &               $\nu$   &      $\Gamma$          \\ \hline
$^\mathrm{P}$P$_1$(1)                      &       3146.96            &     2.15            &       3147.0161(2)         &        4.12(1)        &       3427.29             &        6.46 & 3427.5               &        10(1)       \\ \hline
$^\mathrm{P}$Q$_1$(1)                         &         3165.60          &       2.27          &    3165.6899(2)            &         3.77(2)       &       3445.63              &        6.52 &                 3445.9   &       21(6)       \\ \hline
$^\mathrm{R}$R$_0$(0)                    &         3173.97          &       2.36          &     3174.2373(1)           &         4.66(1)       &           3456.92         &         6.68 &                3457.3&       11.1(3)       \\ \hline
$^\mathrm{R}$R$_1$(1)                        &        3182.61           &         49.92        &      3182.410(6)          &          57.0(6)            &       3469.11              &                  47.11 &       3468.8         &       50(10)        \\ \hline
$^\mathrm{P}$R$_1$(1)                     &       3203.15            &       2.46         &      3203.080(1)          &        8.6(1)        &           --          &          --         &               -- &   --             \\ \hline
\end{tabular}
\end{table}

\end{document}